\documentclass[aps,prl,twocolumn,superscriptaddress,floatfix]{revtex4-1}

\usepackage{graphicx}
\usepackage{float}
\usepackage{amsmath}
\usepackage{bm}

\begin{document}
\setlength{\textfloatsep}{12pt}

\title{Thermal and Nonlinear Dissipative-Soliton Dynamics in \\ Kerr-Microresonator Frequency Combs}


\author{Jordan R. Stone}
\affiliation{Time and Frequency Division, National Institute for Standards and Technology, Boulder, CO 80305}
\email[]{jordan.stone@colorado.edu}
\affiliation{Department of Physics, University of Colorado Boulder, Boulder, CO 80309}
\author{Travis C. Briles}
\affiliation{Time and Frequency Division, National Institute for Standards and Technology, Boulder, CO 80305}
\affiliation{Department of Physics, University of Colorado Boulder, Boulder, CO 80309}
\author{Tara E. Drake}
\affiliation{Time and Frequency Division, National Institute for Standards and Technology, Boulder, CO 80305}
\author{Daryl T. Spencer}
\author{David R. Carlson}
\affiliation{Time and Frequency Division, National Institute for Standards and Technology, Boulder, CO 80305}
\author{Scott A. Diddams}
\affiliation{Time and Frequency Division, National Institute for Standards and Technology, Boulder, CO 80305}
\affiliation{Department of Physics, University of Colorado Boulder, Boulder, CO 80309}
\author{Scott B. Papp}
\affiliation{Time and Frequency Division, National Institute for Standards and Technology, Boulder, CO 80305}
\affiliation{Department of Physics, University of Colorado Boulder, Boulder, CO 80309}


\date{\today}

\begin{abstract}
We explore the dynamical response of dissipative Kerr solitons to changes in pump power and detuning and show how thermal and nonlinear processes couple these parameters to the frequency-comb degrees of freedom. Our experiments are enabled by a Pound-Drever-Hall (PDH) stabilization approach that provides on-demand, radiofrequency control of the frequency comb. PDH locking not only guides Kerr-soliton formation from a cold microresonator, but 
opens a path to decouple the repetition and carrier-envelope-offset frequencies. In particular, we demonstrate phase stabilization of both Kerr-comb degrees-of-freedom to a fractional frequency precision below $10^{-16}$, compatible with optical-timekeeping technology. Moreover, we investigate the fundamental role that residual laser-resonator detuning noise plays in the spectral purity of microwave generation with Kerr combs.
\end{abstract}


\maketitle

\setlength{\parskip}{0em}
Kerr solitons in optical microresonators provide a unique platform for compact, low-noise, microwave-rate, and low-power frequency-comb generation \cite{herr2014temporal,coen2013modeling}. To date, soliton microresonator frequency combs have been used to demonstrate several nonlinear photonics concepts, from soliton crystallization to dark-soliton formation \cite{cole2016soliton, xue2015mode,Yi2016Stokes}, and  micro-scale technologies, including optical clocks \cite{papp2014microresonator}, optical frequency synthesis \cite{spencer2017synth}, communications \cite{pfeifle2014coherent,leo2010temporal,marin2016microresonator}, sensing \cite{suh2016microresonator,Trocha2017Ranging}, and low-noise microwave oscillators \cite{liang2015high}. One central challenge cutting across these directions is the reliable generation of dissipative-Kerr solitons, which are pulses of light balancing nonlinearity, dispersion, gain, and loss. They are parameterized by the relative detuning of the pump laser and Kerr microresonator, and respond to fluctuations in the intracavity field within a few photon lifetimes; as a result, detuning control is critical \cite{yi2016active,guo2016universal,joshi2016thermally,li2017stably,obrzud2016temporal,lucas2017detuning}.

\par Technical issues like bistability \cite{carmon2004dynamical} and mode imperfections \cite{herr2014mode} also impact microresonators and  may suppress soliton formation. Moreover, a fundamental efficiency of Kerr solitons, especially at microwave-rate repetition frequencies, is a high quality factor ($Q$) to enable milliWatt threshold power \cite{ji2017ultra,lee2012chemically}, but this necessitates operation of the pump laser within a narrow, red-detuned frequency window near resonance. Practical experiments utilize servo control to overcome these issues and maintain soliton operation \cite{yi2015soliton,lucas2017detuning}, but this interferes with independent control of the carrier-envelope-offset ($f_{\rm{ceo}}$) \cite{jones2000carrier} and repetition ($f_{\rm{rep}}$) frequencies central to frequency-comb applications. Previous microcomb-locking experiments have leveraged either blue-detuned combs \cite{del2008full}, multiple-soliton states \cite{papp2014microresonator,del2016phase} or lower $Q$ resonators in which laser tunability is less restricted \cite{brasch2017self,briles2017kerr}.

\par In this Letter, we report a general approach to initiate single Kerr solitons from a cold resonator that results in stable radiofrequency (RF) control of the laser detuning, and in turn the soliton dynamics and the frequency comb's $f_{\rm{ceo}}$ and $f_{\rm{rep}}$. Pound-Drever-Hall (PDH) stabilization \cite{drever1983laser} identifies the pumped resonance with high signal-to-noise ratio and locks the laser-resonator detuning to a precise, user-controlled RF frequency. We find that the $f_{\rm{ceo}}$ of Kerr solitons is thermally coupled to the detuning, while the $f_{\rm{rep}}$ dynamics are primarily determined by detuning-dependent Raman scattering. We use our findings to decouple $f_{\rm{ceo}}$ and $f_{\rm{rep}}$ for their straightforward phase stabilization, and to explore low-noise photonic-microwave generation.   


\begin{figure*}[th]
\includegraphics[width = \textwidth]{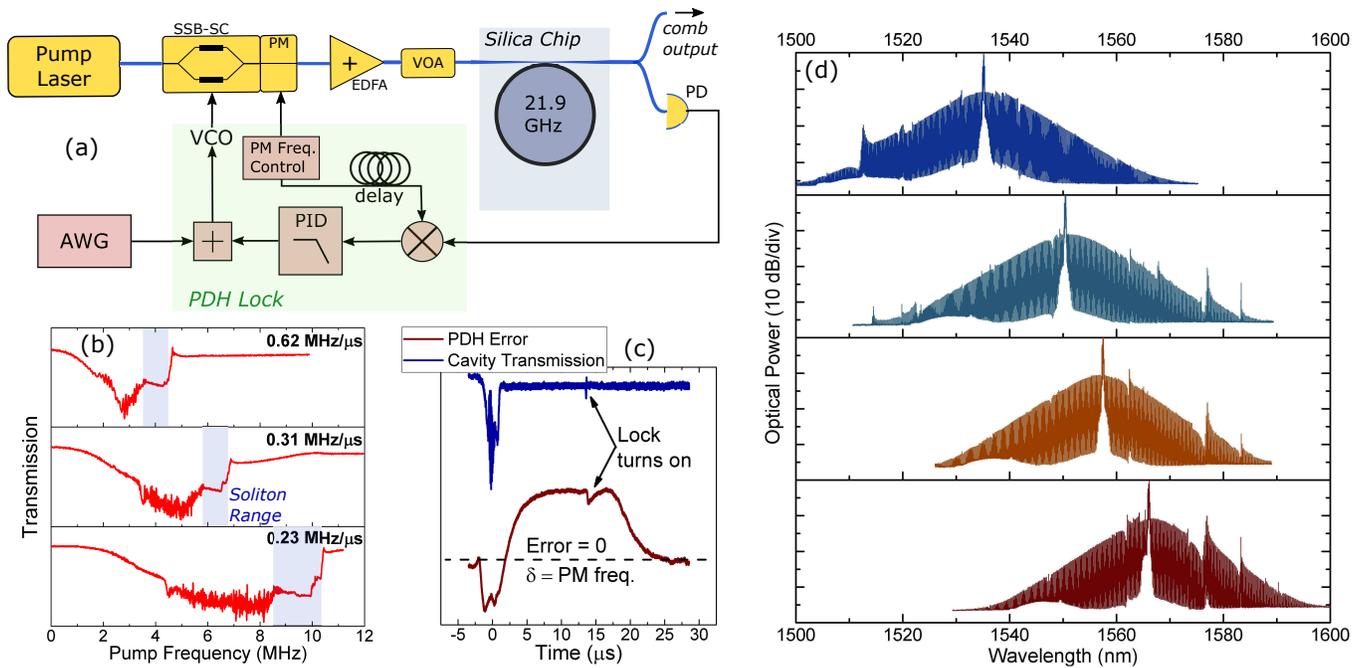}
\caption{\label{Figure1}(a) PDH approach for Kerr-soliton generation. An SSB-SC frequency shifter is driven by a high-bandwidth VCO for fast frequency control of the pump laser, and a servo locks one of the phase-modulation (PM) sidebands at resonance. A voltage-controlled optical attenuator (VOA) is used to control the pump power. (b) By adjusting the frequency sweep rate, we control the transition into the soliton regime. The waveform applied to the VCO is a simple, linear voltage sweep, and the x-axis is relative to the cold cavity resonance frequency. (c) Feedback is initiated at a pre-determined instant of the frequency scan. The dashed line corresponds to the PDH lock point. (d) Generation of soliton frequency combs across the entire C-Band.}
\end{figure*} 

\par We perform the experiments with a 22 GHz free-spectral range (FSR) silica wedge resonator that has a $Q$ of 180 million \cite{yi2015soliton}. The resonator is fabricated on a silicon substrate, and light is coupled to the device via a tapered fiber. Figure \ref{Figure1}(a) presents the experimental setup. The output from an external-cavity diode laser (ECDL) is sent through an SSB-SC frequency shifter composed of a dual-parallel lithium niobate waveguide Mach-Zehnder intensity modulator \cite{wang2015fast} driven by a wideband voltage-controlled oscillator (VCO); we have measured frequency-scan rates up to 100 GHz/$\mu$s with 4 GHz range. By rapidly scanning the pump laser from blue-to-red detuning, we observe the formation of a chaotic Kerr comb followed by the transition to a Kerr soliton; we stabilize the pump laser to the resonance as the soliton waveform settles. To derive the PDH error signal, we apply RF phase-modulation sidebands to the laser before it enters the silica resonator and photodetect them after the resonator. Operationally, the lock point of the PDH servo corresponds to the higher frequency PDH sideband on resonance and the pump laser red-detuned by the phase-modulation frequency \cite{thorpe2008laser,gatti2015wide}. In practice, when a single Kerr soliton is present in the cavity, the PDH sideband probes a cavity resonance weighted towards lower frequencies due to the soliton-induced Kerr shift \cite{guo2016universal}. This introduces a small error between the detuning and phase-modulation frequency, which we estimate to be significantly less than a cavity linewidth (due to the small duty cycle of the soliton pulse train) and therefore do not consider in our experiments. The power in the phase-modulation sidebands is kept $\approx$26 dB below the pump laser, well below the comb-formation threshold, so that the effect of the on-resonance sideband is only to provide a constant thermal shift to the cavity resonance frequency. By precisely adjusting the frequency scan rate to control resonator thermal shifts [Fig. \ref{Figure1}(b)], we optimize so that solitons form in thermal equilibrium. This optimization procedure is thoroughly described in Ref. \cite{briles2017kerr}. We obtain single-Kerr-soliton states across the entire C-band range of the diode laser, as shown in Fig. \ref{Figure1}(d), even as resonator mode-family degeneracies contribute 10 dB excursions to the soliton spectrum. Such flexibility in the pumping frequency is important for applications in frequency synthesis and atomic spectroscopy \cite{spencer2017synth,Stern2017Rb}.

\begin{figure}[t]
\includegraphics[width = \linewidth]{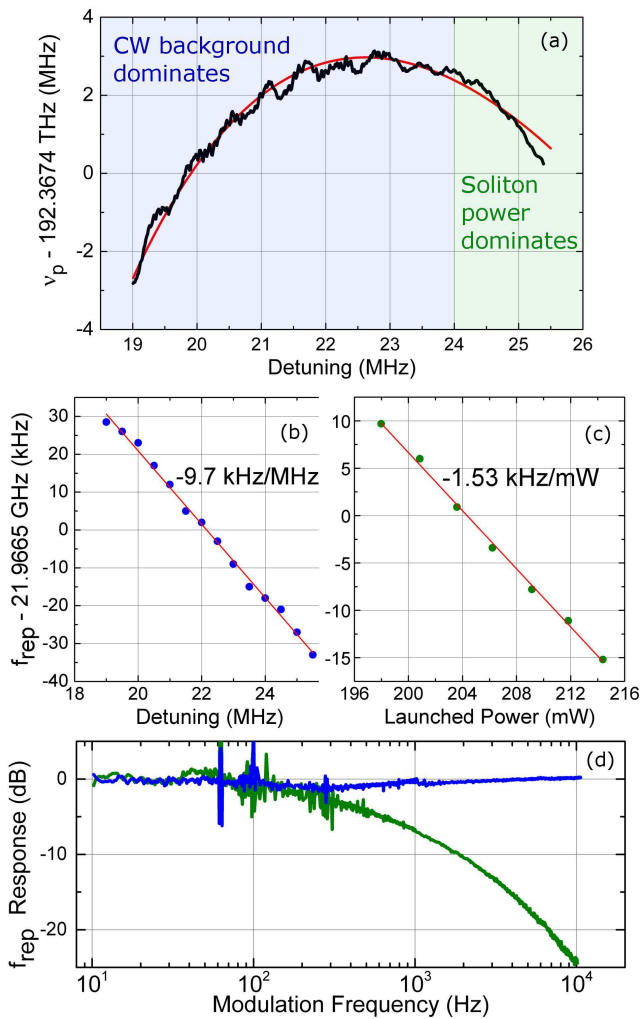}
\caption{\label{Figure2} (a) The CW power and soliton power balance to determine the dependence of pump frequency on detuning. The red curve is the integral of Eq. (\ref{eq:Pcav}) scaled by the cavity frequency rate-of-change with intracavity power. (b) Dependence of soliton repetition frequency with detuning. (c) Dependence of soliton repetition frequency with pump power. Red lines in (b,c) are linear fits. (d) Soliton repetition-frequency response to detuning (blue) and pump-power (green).}
\end{figure}

\par
Building on robust acquisition of single Kerr solitons, we explore their frequency stabilization with respect to an optical reference. This goal, or, alternatively, stabilization through the $f-2f$ technique \cite{diddams2000direct}, requires simultaneous control over both $f_{\rm{rep}}$ and $f_{\rm{ceo}}$, which define the comb-line frequencies through the well-known relation
\vspace{-3pt}\begin{equation}
\nu_{\rm{m}}=f_{\rm{ceo}}+m\times f_{\rm{rep}}, \label{eq:comb}
\end{equation}
where $\nu_m$ is the frequency of each comb mode, indexed by the integer $m$ \cite{jones2000carrier}.
At the same time, the pump-laser frequency $\nu_{\rm{p}}$ serves as a comb line, and Eq. (\ref{eq:comb}) may be rearranged as
\begin{equation}
f_{\rm{ceo}}=\nu_{\rm{p}}-N\times f_{\rm{rep}} \label{eq:offset},
\end{equation}
where $N$ counts the positive integer number of comb modes such that $N\times f_{\rm{rep}} \approx \nu_{\rm{p}}$. Clearly, simultaneous stabilization of $\nu_{\rm{p}}$ and $f_{\rm{rep}}$ implies the stability of $f_{\rm{ceo}}$, which is the relevant parameter for users of the comb who likely do not require (or desire) knowledge of the Kerr-comb dynamics internal to the microresonator. Therefore, $\nu_{\rm{p}}$ must be adjusted to serve two purposes at the same time: (1) Maintain $\delta$ within the appropriate range for stable Kerr-soliton propagation and (2) Be phase-locked or tuned subject to user requirements. 

\par Satisfying the above criteria for $\nu_{\rm{p}}$ while maintaining independent control over $f_{\rm{rep}}$ necessitates the decoupling of these two degrees-of-freedom. We therefore search for settings of pump power and detuning that allow $f_{\rm{rep}}$ to be adjusted without disturbing $\nu_{\rm{p}}$. This leads us to study the thermal and nonlinear processes that couple $\nu_{\rm{p}}$ and $f_{\rm{rep}}$; in particular, our investigation allows us to map the response of all comb-line frequencies to changes in pump power and detuning (Fig. \ref{Figure2}). To start, we recall that the repetition frequency of a soliton Kerr comb is approximated by 
\begin{equation}
2\pi f_{\rm{rep}}=D_1+\Omega \, D_2/D_1 \label{eq:rep-rate},
\end{equation}
where $D_1$ is the FSR in radians per second, $D_2$ is the second-order dispersion about the pump frequency \cite{herr2012universal}, and $\Omega$ is the soliton self-frequency shift (SSFS) that results from a combination of Raman and mode-perturbation effects \cite{yi2017single}. The SSFS describes a frequency shift in the comb spectrum relative to the pump frequency, and is therefore coupled to the repetition rate through second-order dispersion. Moreover, the SSFS is generally dominated by the Raman nonlinearity, which produces an SSFS linear in $\delta$ \cite{yi2017single,karpov2016raman}. To determine how this couples $f_{\rm{rep}}$ to $\nu_{\rm{p}}$, we analyze the detuning dependence of the latter. 


Central to our study is our finding that $\nu_{\rm{p}}$ does not depend linearly on $\delta$ (since we control $\delta$ directly, it is necessary to think of $\nu_{\rm{p}}$ as the dependent variable, though the dynamics we outline here also apply to a free-running pump laser). Specifically, we find that some settings of pump power and detuning enable the decoupling of $\nu_{\rm{p}}$ from $\delta$. This surprising result may be understood by considering the interplay between intracavity power ($P_{\rm{cav}}$), $\delta$, and the cavity resonance frequency. Changes in $P_{\rm{cav}}$ will modify the microresonator temperature, changing its index of refraction (and thus its mode spectrum) via the thermo-optic effect \cite{carmon2004dynamical}. For a single circulating Kerr soliton, the intracavity field is comprised of both the soliton and a continuous-wave (CW) background associated with the pump laser. Since both of these contribute to the total optical power, the rate-of-change of $P_{\rm{cav}}$ with $\delta$, in the regime $\delta\gg\Gamma$ ($\Gamma$ is the resonator linewidth, approximately 1.1 MHz in our system), is \cite{yi2015soliton,herr2014temporal,matsko2013timing}

\begin{equation}
\frac{\partial P_{\rm{cav}}}{\partial \delta}=\frac{n\, A_{\rm{eff}}}{2\pi n_2 \nu_c}\sqrt{\frac{2D_2}{2\pi \delta}} - \eta\frac{\mathcal{F}}{\pi}\frac{\Gamma^2P_{\rm{in}}}{\delta^3}, \label{eq:Pcav}
\end{equation} 
where $n$ is the refractive index, $A_{\rm{eff}}$ is the effective mode area, $n_2$ is the Kerr index, $\nu_{\rm{c}}$ is the frequency of the pumped mode, $\eta$ is the coupling efficiency, $\mathcal{F}$ is the cavity finesse, and $P_{\rm{in}}$ is the pump power. According to Eq. (\ref{eq:Pcav}), the soliton pulse and CW background [the first and second terms on the right side of Eq. (\ref{eq:Pcav}), respectively] compete to determine the sign of $\partial P_{\rm{cav}}/\partial \delta$, and which term dominates depends on the magnitude of $\delta$. While the CW background primarily determines $\partial P_{\rm{cav}}/\partial \delta$ at small $\delta$, it becomes negligible at larger $\delta$, and Eq. (\ref{eq:Pcav}) may be approximated using only the soliton term. Hence, we find that Kerr-soliton frequency combs operate in essentially two $\delta$ regimes: (1) $\partial P_{\rm{cav}}/\partial \delta < 0$ and $\nu_{\rm{p}}$ tunes opposite to $\delta$, (2) $\partial P_{\rm{cav}}/\partial \delta > 0$ and $\nu_{\rm{p}}$ tunes with the same sign as $\delta$. At the crossover of these two regimes, small changes in $\delta$ transfer power evenly between the soliton and CW background and the cavity becomes thermally decoupled from $\delta$. Near this point (see Fig. \ref{Figure2}a), where changes in $\delta$ are offset by the thermal shifts they induce in the resonance frequency, $\nu_{\rm{p}}$ corresponds to a ``fixed point'' of the frequency comb \cite{telle2002kerr}.


To test our understanding of these dynamics, we vary $\delta$ through the PDH lock and record changes in the pump-laser frequency [Fig. \ref{Figure2}(a)]. For comparison, we integrate Eq. (4) and multiply by a measured cavity tuning coefficient of ~50 MHz/W to generate a prediction curve for Fig. \ref{Figure2}(a). Values for Eq. (4) parameters are: $A_{\rm{eff}}$ = 60 $\mu \rm{m}^2$ \cite{yi2015soliton}; $n_2$ = $2.6\times10^{-20}$ $\rm{m}^2$/W; $D_2/2\pi$ = 14 kHz; $\mathcal{F}$ = 20,000; $P_{\rm{in}}$ = 250 mW. The coupling parameter $\eta$ is used as a fitting parameter and allowed to vary around 0.7, chosen because the system is slightly overcoupled to improve efficiency \cite{yi2015soliton}. We find $\eta$ = 0.62 fits the data well. Equation (4) accurately predicts the $\nu_{\rm{p}}$ behavior in both the blue-shaded region, where the pump dynamics are largely determined by the CW background, and the green region, where the soliton physics dominates. In particular, the model identifies the turning point between 22 MHz and 23 MHz. Additionally, we record in Fig. \ref{Figure2}(b,c) the dependence of $f_{\rm{rep}}$ on both $\delta$ and pump power. Evidently, either of these may be used to tune the repetition frequency; however, the response of $f_{\rm{rep}}$ is different in the two cases. Because the pump power relies on thermal effects to control $f_{\rm{rep}}$, the response bandwidth is limited by the resonator thermal response time \cite{carmon2004dynamical}, whereas control of $f_{\rm{rep}}$ through $\delta$ is limited by the SSFS response and practically limited by the bandwidth of the PDH lock [see Fig. \ref{Figure2}(d)] \cite{loh2016microrod}.

\begin{figure}[t]
\includegraphics[width=\linewidth]{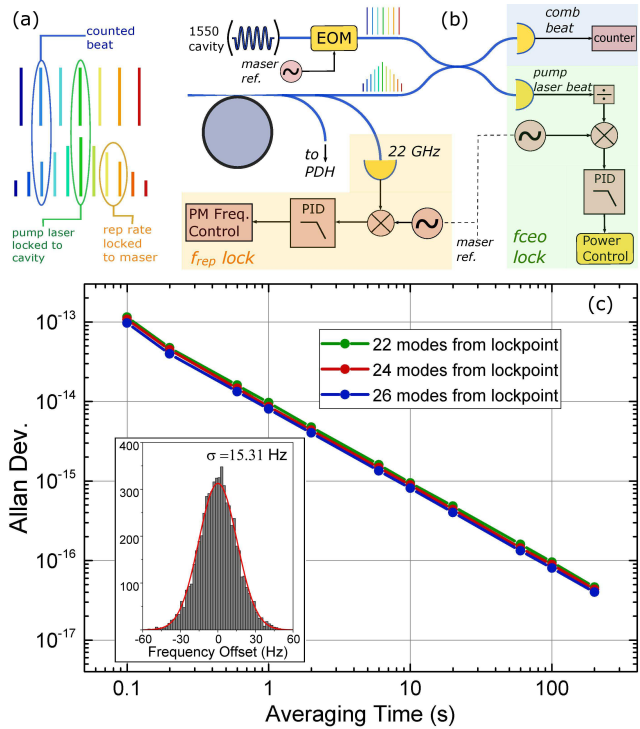}
\caption{\label{Figure3} (a) Illustration of the experiment to characterize residual noise in our phase locks of $f_{\rm{rep}}$ and $f_{\rm{ceo}}$. The pump laser is phase-locked to a cavity-stabilized laser, which is electro-optically (EO) modulated to produce a reference comb. The repetition rates of both combs are derived from the same reference. (b) Setup used to fully stabilize the Kerr comb. (c) Allan deviations of three Kerr comb lines counted against neighboring EO teeth. Inset: Distribution of counted frequencies at 0.1 s gate time with a Gaussian fit (red line). The error between the mean and the expected value is 2 mHz.}
\end{figure}

In view of the results shown in Fig. \ref{Figure2}, an optimal stabilization strategy is to decouple the frequency-comb degrees-of-freedom by operating about the detuning that corresponds to a $\nu_{\rm{p}}$ fixed point. Figure 3 presents a detailed schematic and measurements that demonstrate this strategy. We tune $f_{\rm{rep}}$ through feedback to $\delta$ (by modulating the PDH frequency) and phase-lock it to a hydrogen-maser-referenced $\sim$22 GHz oscillator. We directly phase-lock $\nu_{\rm{p}}$ to an ultralow-expansion glass Fabry-Perot (FP) stabilized laser at 1550 nm, actuating the Kerr comb's pump power. To characterize the residual noise in our optical and microwave phase-locks, we form an electro-optic (EO) frequency comb around the FP-stabilized laser, using a microwave oscillator synthesized from the same H-maser reference; see Fig. \ref{Figure3}(a). The stability of optical-heterodyne beatnotes between the Kerr and EO combs quantifies the residual-frequency-noise of our two phase locks. Allan deviation measurements \cite{riehle2006frequency} are shown in Fig. \ref{Figure3}(c). The performance of our Kerr-comb system, stable to within $10^{-16}$ imprecision, enables a compact platform for frequency metrology and is commensurate with modern optical-timekeeping technology. 

\begin{figure}[t]
\includegraphics[width=\linewidth]{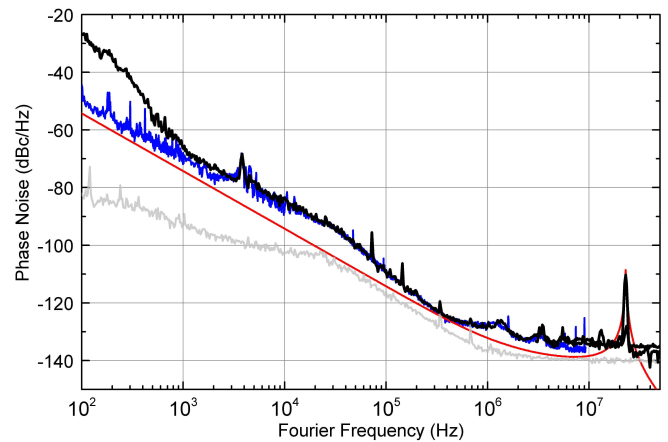}
\caption{\label{Figure4} Measured phase noise of $f_{\rm{rep}}$ (black traces), and predictions obtained from our model (red) and the PDH error signal (blue). The gray trace is the measurement-system floor.}
\end{figure}

\par In addition to stable detuning control, our PDH scheme provides a unique opportunity to study the transduction of detuning fluctuations into frequency-comb noise, since the fluctuations manifest as residual noise in the PDH error signal. We explore the phase noise $L_{\rm{\phi}}(f)$ of our Kerr comb's 21.98 GHz repetition frequency when the detuning is stable but the system is otherwise free-running. Since $L_{\rm{\phi}}(f)$ is lower than that of many commercial microwave synthesizers, we use a self-referenced EO frequency comb operated as an optical-frequency divider to provide a reference oscillator at 22 GHz \cite{Carlson2017EO}. Prior to photodetecting $f_{\rm{rep}}$, we bandstop filter residual pump light, and re-amplify the remaining soliton comb from $\sim$300 $\mu$W to $\sim$10 mW. Our measurement is shown in Fig. \ref{Figure4}. Of particular interest is the high-Fourier-frequency noise, which is significantly above both the shot-noise level ($\approx$-160 dBc/Hz accounting for amplifier noise figures) and the measurement floor. Understanding this issue is important for future applications. For instance, in experiments relying on the spectral broadening of Kerr solitons, the high-frequency noise plays a key role \cite{lamb2017optical}.

\par 
After calibrating the PDH error signal, we record its Fourier spectrum and multiply by the transfer function in Fig. \ref{Figure2}(d). The resulting spectrum (blue trace in Fig. \ref{Figure4}) gives the expected contribution of detuning noise to $L_{\rm{\phi}}(f)$. Since this curve reproduces our $L_{\rm{\phi}}(f)$ measurement well for Fourier frequencies outside the thermal bandwidth, we conclude that detuning noise is the most important contribution to the microwave spectral purity for Kerr solitons exhibiting a large SSFS. Separately, we model the contributions to $L_{\rm{\phi}}(f)$ by analyzing $\nu_{\rm{p}}$-to-$f_{\rm{rep}}$ noise conversion. With $\delta\gg\Gamma$, the resonator selectively enhances the typical white-frequency-noise spectrum of an ECDL at the Fourier frequency $f=\delta$. Specifically, we predict that
\begin{equation}
L_{\rm{\phi}}(f)\approx\frac{1}{f^2}\left(\frac{\partial f_{\rm{rep}}}{\partial \delta}\frac{\nu_{\rm{c}}n_2}{nA_{\rm{eff}}}\right)^2S_{\rm{I, cav}}(f), \label{eq:NoiseModel}
\end{equation}
where $\partial f_{\rm{rep}}/\partial \delta$ is the conversion factor of $\delta$ noise to $f_{\rm{rep}}$ frequency noise, and $S_{\rm{I, cav}}(f)$ is the intracavity intensity noise calculated for a white-frequency-noise pump laser \cite{kippenberg2013phase}. In Eq. (\ref{eq:NoiseModel}), the Kerr nonlinearity converts $S_{\rm{I, cav}}(f)$ into detuning fluctuations that couple to $L_{\rm{\phi}}(f)$ through $\partial f_{\rm{rep}}/\partial \delta$ (see supplemental material for more details). The red trace in Fig. 4 shows how this model mostly captures our measured $L_{\rm{\phi}}(f)$ noise floor. Thus, a lower-noise pump laser (or a higher bandwidth PDH lock) should dramatically improve the microwave spectral purity; this prediction is confirmed experimentally in Ref. \cite{lamb2017optical}.

\par In summary, we have introduced a novel Pound-Drever-Hall system for generating, studying, and controlling dissipative-Kerr solitons in microresonators. We already utilize the technique with multiple microresonator platforms \cite{briles2017kerr,lamb2017optical}, including in SiN resonators that had previously required a specific dispersion profile to balance the thermal bistability and mode-perturbation effects \cite{li2017stably}. Rapid frequency scanning and PDH locking could be implemented with discrete semiconductor lasers and Kerr microresonators, or potentially in a heterogeneously integrated Kerr-comb platform. 

\par We thank X. Yi, K. Yang, and K. Vahala at Caltech for assisting and for fabricating the resonator used in the experiment, as well as D. Hickstein, L. Stern, Z. Newman, and F. Quinlan for valuable comments on the manuscript. This project was funded by the DARPA DODOS program, AFOSR under award number FA9550-16-1-0016, and NIST. This work is a contribution of the U.S. government and is not subject to copyright.

\bibliography{Draft1_LaTex}

\end{document}